\documentclass[prb,preprintnumbers,twocolumn,amsmath,amssymb,floatfix]{revtex4-1}

\usepackage{graphicx}% Include figure files
\usepackage{amsmath}
\usepackage{amssymb}
\usepackage[normalem]{ulem}
\usepackage{color}
\usepackage[spacing=true,factor=1100,stretch=10,shrink=10]{microtype}
\microtypecontext{spacing=nonfrench}

\newcommand{\hcd}{\hat{c}^{\dagger}}

\newcommand{\hc}{\hat{c}^{\phantom{\dagger}}}
\newcommand{\comments}[1]{}

\begin{document}

\title{Band antiferromagnetism in a two-band model for iron pnictides}
\author{Tobias Schickling}
\affiliation{Fachbereich Physik, Philipps Universit\"at Marburg, Renthof 6, 35032
Marburg, Germany}

\begin{abstract}
	In this work I investigate a two-band Hubbard model using
	the Gutzwiller wavefunction. The tight-binding part of the model was
	constructed to have a gapless spin-density wave state which leads to 
	Dirac points in the bandstructure, a common feature of
	many iron-pnictide compounds.
	For quarter, half and three-quarter fillings I show that the Hund's rule
	coupling has a large impact on the metal-insulator transition in the
	paramagnetic phase. 
	For the half-filled model in the antiferromagnetic phase,
	the magnetism evolves in a Stoner-like behavior and 
	the size of the ordered moment is mainly determined by the Hubbard
	interaction.
	As the Hund's coupling plays a minor role in this state, the model
	does not describe a Hund's metal which is in contrast to more realistic
	models for iron-pnictide compounds.
\end{abstract}

\maketitle

%-----------------------------------------------------------------------
%	Introduction
%-----------------------------------------------------------------------

The study of (antiferro)magnetism in multi-band Hubbard models has received a
strong push by the recently discovered iron-based superconductors.  The ground states of
many of these materials are characterized by an antiferromagnetic state (AFM) with a
small ordered moment. This small magnetic moment cannot be captured by Density
Functional Theory (DFT)
calculations.~\cite{Mazin:2008a, Singh:2008, Yildirim:2008}

To understand the discrepancy between the predictions of DFT and experiments,
electron-electron correlations are believed to play an important role. This
issue was addressed by several groups employing the Dynamical Mean-Field Theory
(DMFT).~\cite{Ishida:2010, Aichhorn:2011, Yin:2011_b} They have shown that the ground state of a multi-band Hubbard
model with realistic values of $(U,J)$ is an ordered \textit{stripe} AFM
(metallic), with a magnetic moment that is strongly reduced with respect to the
LSDA value ($m_{\mathrm{LDA}} \simeq 2.0 \mu_B$). Especially, the ordered magnetic moment is
significantly smaller than the local moment. This particular
ground state was named a \textit{Hund}'s metal, to signify that in these correlated
multi-band metals, the Hund's coupling plays a more important role than the
on-site Coulomb correlation $U$.~\cite{Yin:2011_b}

The same questions were recently also studied by employing the Gutzwiller wave
function.~\cite{Yao:2011_a, Schickling:2011,Schickling:2012}
It was shown that the Gutzwiller approximation captures the correct physics,
\textit{i.e.}, a small ordered magnetic moment and a large local moment. Additionally,
the small ordered moment phase reveals a very simple
Stoner (Slater) picture, \textit{i.e.}, the value of the magnetic moment is
essentially
determined by \textit{band} physics. The ground state displays the
\textit{band magnetism of correlated quasi-particles}, \textit{i.e.}, an
intermediate state between the local moment (Heisenberg) picture and the
itinerant (Slater) picture. Both pictures have been proposed as a
starting point for the magnetic ground state of iron pnictides.~\cite{Mazin:2008c,
Si:2008, Stanek:2011}

Besides magnetism the explanation of superconductivity in iron-pnictide
compounds remains an important issue. Unfortunately, the rich band structure of
realistic models complicates the discussion. Therefore, one tries to reduce the
complexity of the models, and a minimal model for the iron pnictides usually
contains two bands with $d_{xz}$ and $d_{yz}$
orbitals.~\cite{Hu:2012,Lo:2013,Tai:2013,Quan:2012}

In this study, I address the question how many-particle correlations
influence the magnetic properties and the band structure of such a two-band
model by using the Gutzwiller wavefunction.  In the first part of this work I
investigate the paramagnetic phase of the model. I show that
the metal-insulator transition depends strongly on the size of the Hund's
rule coupling. In the second part of this work I show how
antiferromagnetism evolves in the half-filled model. The magnetic moment
is determined by an effective energy scale that depends mainly
on the Hubbard interaction, and the Hund's rule exchange has only a small
impact.  Therefore, the investigated two-band model can not be considered as
a Hund's metal which is assumed to be an important property of
iron-pnictide compounds.

\begin{figure*}
	\includegraphics[width=\textwidth]{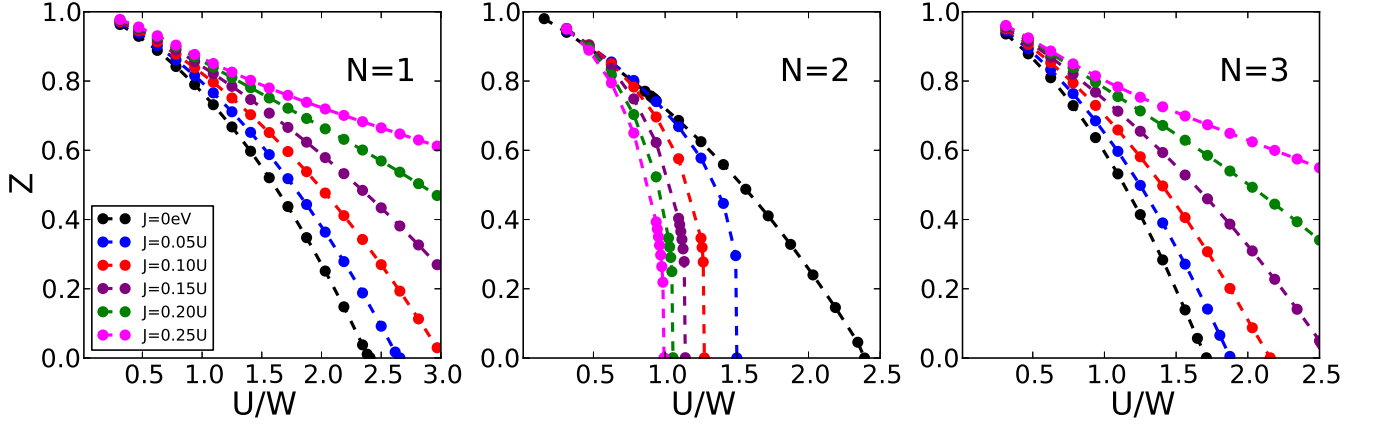}
	\caption{Quasi-particle renormalization for the two-band model as a function of the
	Hubbard interaction $U$ for various $J/U$ and for the electron numbers
	$N=1,2,3$. The quantitative different results for $N=1$ and $N=3$ reflect the
	fact that the model is not particle-hole symmetric.}
	\label{fig:gwpara}
\end{figure*}

The investigated two-band model was introduced by Ran
\textit{et al.},~and contains essential topological single-particle properties of
a large group of the iron pnictides.~\cite{Ran:2009}  In a previous publication,
the model was studied using a slave-rotor approach which leads to a rich phase
diagram.~\cite{Ko:2011} In the current study I apply the Gutzwiller wavefunction
to the model.

Compared to most other methods, the Gutzwiller ansatz is
numerically fairly cheap. Therefore, it is possible to scan a large part of the
$(U,J)$ parameter space as compared to other methods.~\cite{Buenemann_2012_b}

%-----------------------------------------------------------------------
%	Model and method
%-----------------------------------------------------------------------
%TB Hamiltonian
I investigate the following model in two dimensions,
\begin{equation}\label{equ:fullHamiltonian}
	\hat{H} = \hat{H}_0 + \hat{H}_{C} = \displaystyle\sum_{i,j;b,b^{\prime};\sigma}
				t_{i,j}^{b,b^{\prime}} \hat{c}^{\dagger}_{i,b,\sigma}
				\hat{c}_{j,b^{\prime},\sigma} +\displaystyle\sum_{i} \hat{H}_{C,i},
\end{equation}
where $\hat{H}_0$ describes the kinetic energy of the electrons and
$\hat{H}_{C}$ includes the correlation part. As already mentioned, for the
kinetic energy term, I
employ the two-band model of Ran \textit{et al.}~\cite{Ran:2009} The underlying
lattice is quadratic, with $x$ and $y$ axes directed along the edges of the square.
The two orbitals have $XZ$ and $YZ$ symmetry and the total bandwidth 
of the model is $W = 12.8\,\mathrm{eV}$. For the correlated part
of the Hamiltonian, I choose
{\arraycolsep=1pt\begin{eqnarray}
\hat{H}_{\rm C}&=&\hat{H}^{\rm dens}_{\rm C}+\hat{H}^{\rm sf}_{\rm C} \; ,
\nonumber\\[3pt]
\hat{H}^{\rm dens}_{\rm C}&=&
U \displaystyle\sum_{b}
\hat{n}_{b,\uparrow}\hat{n}_{b,\downarrow}
+\!\!
\displaystyle\sum_{\scriptstyle \sigma,\sigma'}
\widetilde{U}_{\scriptstyle \sigma,\sigma'}
\hat{n}_{XZ,\sigma}\hat{n}_{YZ,\sigma'} \, ,\nonumber\\
\hat{H}^{\rm sf}_{\rm C} &=&
J\sum_{\sigma}\hcd_{XZ,\sigma}\hcd_{YZ,\bar{\sigma}}
\hc_{XZ,\bar{\sigma}}\hc_{YZ,\sigma}
\label{app3.5}\\
&&+J\left(\hcd_{XZ,\uparrow}\hcd_{XZ,\downarrow}
\hc_{YZ,\downarrow}\hc_{YZ,\uparrow}+ {\rm h.c.}\right)  \;.
\nonumber
\end{eqnarray}}%
Here, I dropped the lattice-site indices and introduced the abbreviations
$\bar{\uparrow}=\downarrow$, $\bar{\downarrow}=\uparrow$, and
$\widetilde{U}_{\sigma,\sigma'}=(U^{\prime}-\delta_{\sigma,\sigma'}J)$, where $U$, $U^{\prime}$
and  $J$ are the local Coulomb and exchange interactions. For $t_{2g}$-orbitals
the relation $U=U^{\prime}+2J$ holds.
In the current study an equi-spaced grid of $U,J$ values, for $U < 20.0\,\mathrm{eV}$ and
$J < 2.0\,\mathrm{eV}$ is explored. To investigate the saturation of the magnetic moment, I push $U$
and $J$ up to $30\,\mathrm{eV}$ and $5\,\mathrm{eV}$ respectively for selected
isocontours. 

%Method
The Gutzwiller variational wave function
\begin{equation}
|\Psi_{\rm G}\rangle=\hat{P}_{\rm G}|\Psi_0\rangle
=\prod\nolimits_{i}\hat{P}_{i}|\Psi_0\rangle \label{1.3} \end{equation}
approximates the true ground state of $\hat{H}$ in~(\ref{equ:fullHamiltonian}).
The wave function $|\Psi_0\rangle$ is a product state of filled Bloch orbitals
and is determined self-consistently. The operator $\hat{P}_{i}$ is the local
Gutzwiller correlator which is defined as 
\begin{equation}
	\hat{P}_{i}=\sum\nolimits_{\Gamma}\lambda_{\Gamma} |\Gamma \rangle_{i}
	{}_{i}\langle \Gamma |\;.  
	\label{1.4} 
\end{equation}
Here, $|\Gamma \rangle_{i}$ are the eigenstates of $\hat{H}_{{\rm C},i}$,
and for each eigenstate variational
parameters $\lambda_{\Gamma}$ are introduced.
The wave function~(\ref{1.3}) generates the energy functional
$E[|\Psi_G\rangle] = \langle \hat{H} \rangle_{\Psi_G}$ that
has to be minimized with respect to the single-particle product state
$|\Psi_0\rangle$ and the parameters $\lambda_{\Gamma}$.

In the limit of infinite spatial dimensions, expectation values can
be evaluated without further approximations.~\cite{Buenemann:1998, Buenemann:2005}
For finite-dimensional systems, I use this energy functional as an
approximation ('Gutzwiller approximation'). This 
approach also provides the Landau--Gutzwiller quasi-particle bandstructure
that can be compared with ARPES data.~\cite{Buenemann:2003, Hofmann:2009,
Buenemann_2012_b}

I begin the discussion with considering the paramagnetic state of the
system with average electron numbers $N=1,2,3$ per site which corresponds to quarter, half 
and three-quarter fillings. The results of the Gutzwiller
calculations are shown in Fig.~\ref{fig:gwpara}. For these calculations,
I fixed the $J/U$-ratio and Fig.~\ref{fig:gwpara} shows the
quasi-particle renormalization $Z$ as a function of the Hubbard $U$. 
For $Z$ going to zero, I find a metal
insulator transition (MIT) of the Mott-Hubbard type (Brinkman-Rice
transition~\cite{Brinkman:1970}). The critical value of the Hubbard
interaction for this MIT, $U_c$, is
typically not well estimated in the Gutzwiller approximation.
Nonetheless, the \textit{qualitative} trend is usually
correct. Therefore, the results shown in Fig.~\ref{fig:gwpara} 
are consistent with a recent DMFT model study.~\cite{deMedici:2011_a, deMedici:2011_b}

%%%%%%
%% Figure 2: Fermi surface
%%%%%%
\begin{figure}[ht]
	\centering
	\begin{minipage}{0.19\textwidth}
		\includegraphics[width=\textwidth]{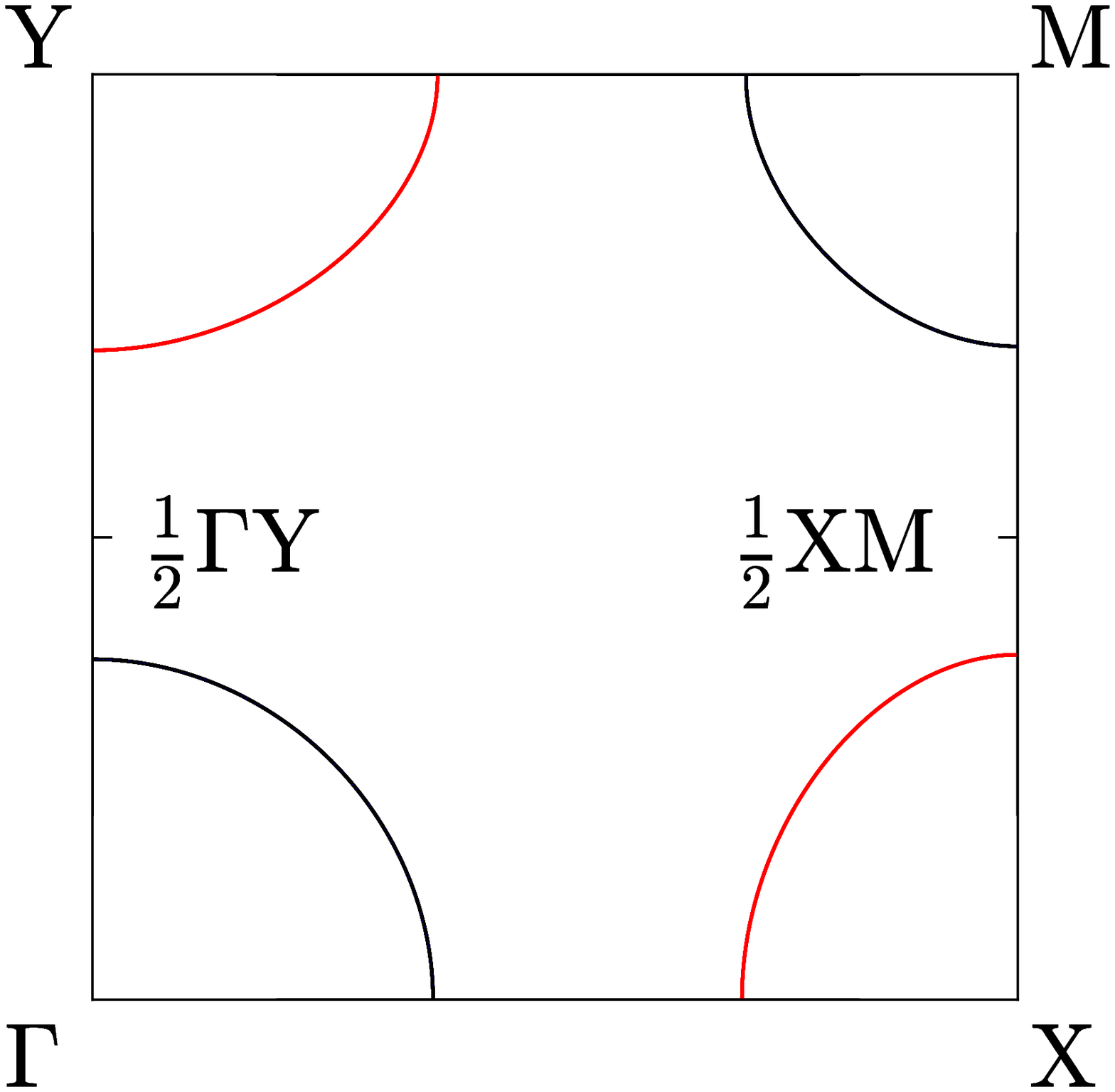}

		\centering{(a)}
	\end{minipage}
	\hspace{0.02\textwidth}
	\begin{minipage}{0.24\textwidth}
		\includegraphics[width=\textwidth]{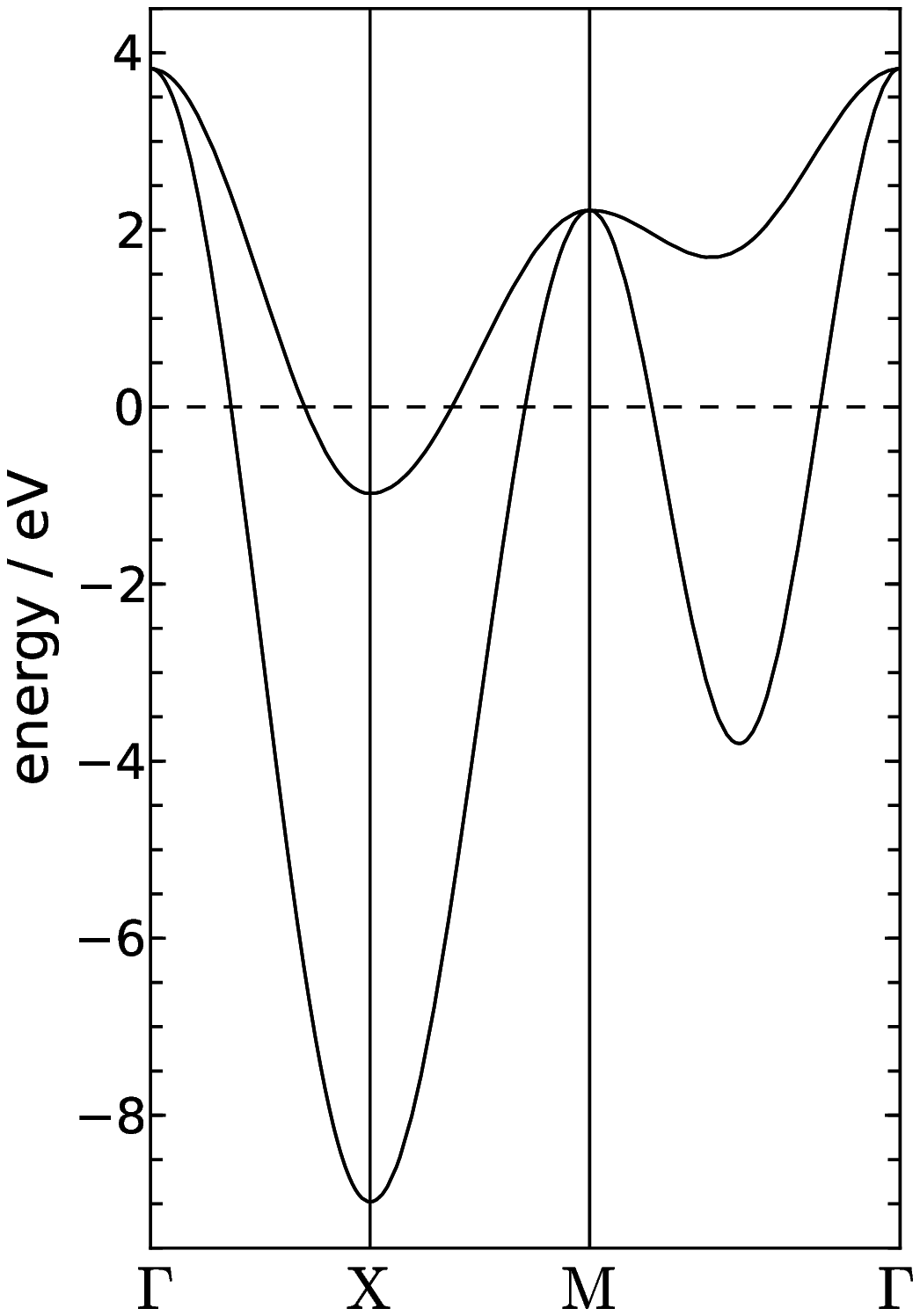} 
		
		\centering{(b)}
	\end{minipage}
	\caption{
		Properties of the bare two-band model: 
		(a) Fermi surface with hole pockets around
		the $\Gamma$ and the $\mathrm{M}$ point (black) and electron
		pockets around the $\mathrm{X}$ and the $\mathrm{Y}$ point (red);
		(b) bare bandstructure.}
	\label{fig:bandsBare} 
\end{figure}

For quarter filling and $J=0$, the quasi-particle weight goes 
to zero at $U_c$ $\sim 2.5\,W$. 
The critical value for the Hubbard interaction increases with $J/U$.
For $N=1$, a larger Hund's rule coupling suppresses the MIT.
If the system is half filled, \textit{i.e.}~$N=2$, the Hund's rule coupling
has the opposite effect. For $N=2$ and $J=0$, the critical value for
the Hubbard interaction is also $U_c$ $\sim 2.5\,W$. However, the critical
value decreases with increasing $J$, \textit{e.g.},~for $J/U=0.15$ the 
critical value is reduced to $\sim
1.2\,W$. The case of a three quarter filling of the model, $N=3$,
is similar to quarter filling, \textit{i.e.}~$N=1$. For $N=3$ and $J=0$, the MIT 
occurs at $U_c\sim 1.6\,W$, \textit{i.e.},~the system with
electron number $N=3$ is more correlated than the system with a smaller number of
electrons. For a finite $J$, the critical $U$-value increases as a function of $J$,
as in the quarter filled case, $N=1$. The explanation of this observation can be
related to the fact that the system is not particle-hole symmetric. Therefore,
the trends for $U_c(J)$ are the same for $N=1$ and $N=3$ although there are
quantitative differences.
Similar results have been observed in a previous DMFT study.~\cite{deMedici:2011_a}
This is especially interesting since the model from the DMFT study contains three
orbitals and is particle-hole symmetric. 

To summarize these results, in the paramagnetic case the system shows a strong
dependency on the size of the Hund's rule coupling. As already observed in
previous studies, the Hund's rule coupling
decreases ($N=1,3$) or increases ($N=2$) correlation effects, depending on the
number of electrons. 
In contrast, the antiferromagnetic phase of the half-filled two-band model only
weakly depends on $J$, as I will show next.

The present two-band model may be considered a minimal model for iron-pnictide
superconductors with emphasis on the topological aspects of the bandstructure.~\cite{Ran:2009}
In the following discussion of the antiferromagnetic ground state, I focus on the
system at half filling and the corresponding Fermi surface of the model is
shown in Fig.~\ref{fig:bandsBare}.
The Fermi surface  -- Fig.~\ref{fig:bandsBare}\,$(a)$ --
comprises hole pockets around the $\Gamma$ and the $M$ point, and two electron pockets around
the $X$ and the $Y$ point, with $XZ$ and $YZ$ orbital characters. 
This Fermi surface topology shows some similarities in comparison with the Fermi surface
topology of more realistic models for iron-based superconductors, \textit{e.g.},~an
eight-band model.~\cite{Andersen2011}
The main differences are that the hole pocket at the $\Gamma$ point is much smaller in
the more realistic model and, as a result of the more complex bandstructure, 
two hole pockets are located at the $M$ point.

Nevertheless, in accordance with more realistic models the Fermi surface topology 
of the present two-band model leads to a strong
nesting between the \textit{hole} and \textit{electron} pockets with vector
$\mathbf{Q}=(0,\pi)$ (the vector $\mathbf{Q}=(\pi,0)$ is equivalent, but was not
used in this work).  This corresponds in real space to a \textit{stripe} 
spin-density wave (SDW) pattern, in which the spins align ferromagnetically along one
of the two equivalent directions ($x$ in the following), and
antiferromagnetically along the other ($y$).  In this SDW state, the bands fold
on top of each other along the 1/2 $\Gamma\mathrm{Y}$-1/2 $\mathrm{XM}$ line,
bringing the $Y$ point on top of the $\Gamma$ point, and the $M$ point on top of
the $X$ point.

\begin{figure}[htb]
	\centering
	\begin{minipage}{0.22\textwidth}
		\includegraphics[width=\textwidth]{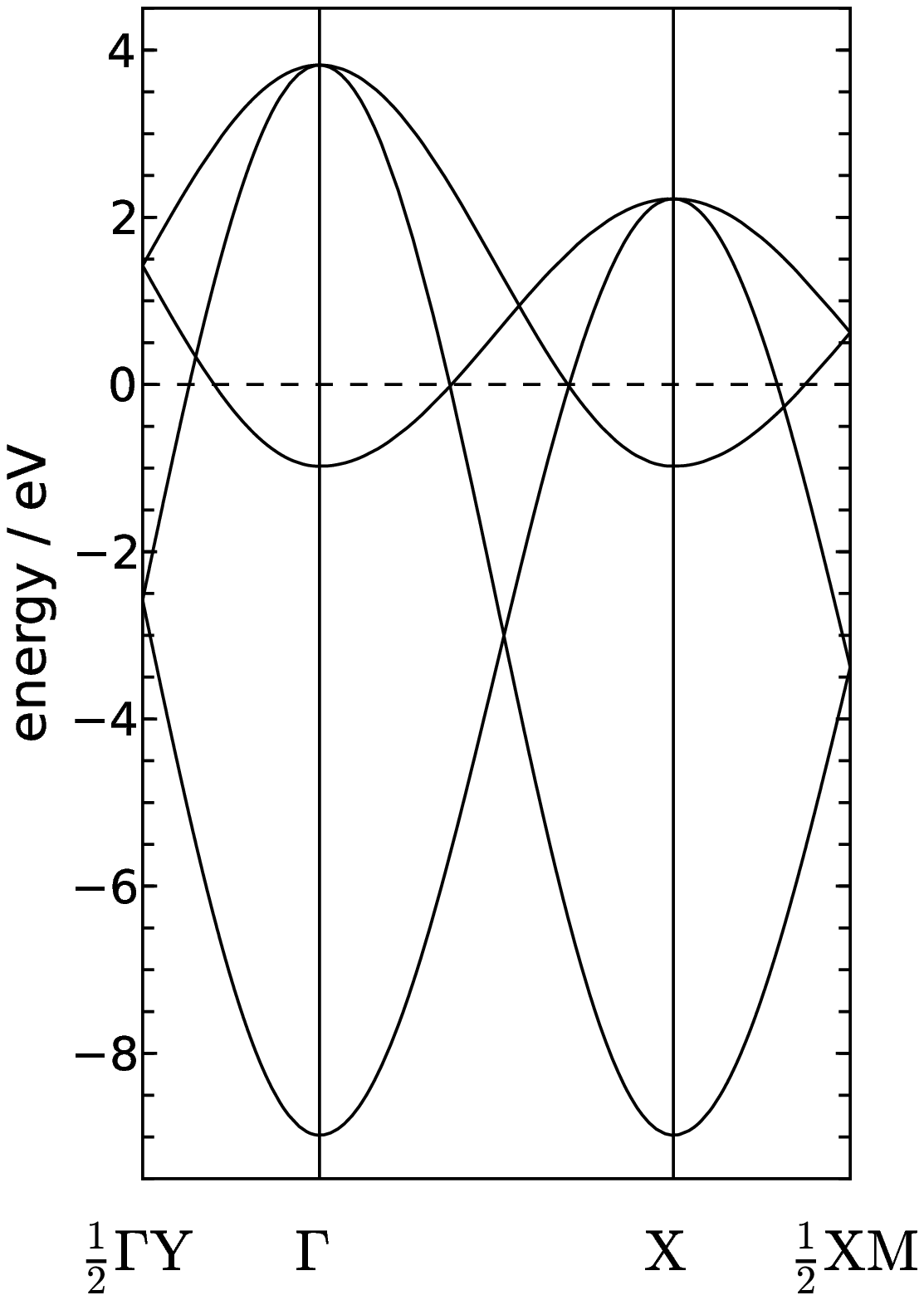} 

		\centering{(a)}
	\end{minipage}
	\hspace{0.02\textwidth}
	\begin{minipage}{0.22\textwidth}
		\includegraphics[width=\textwidth]{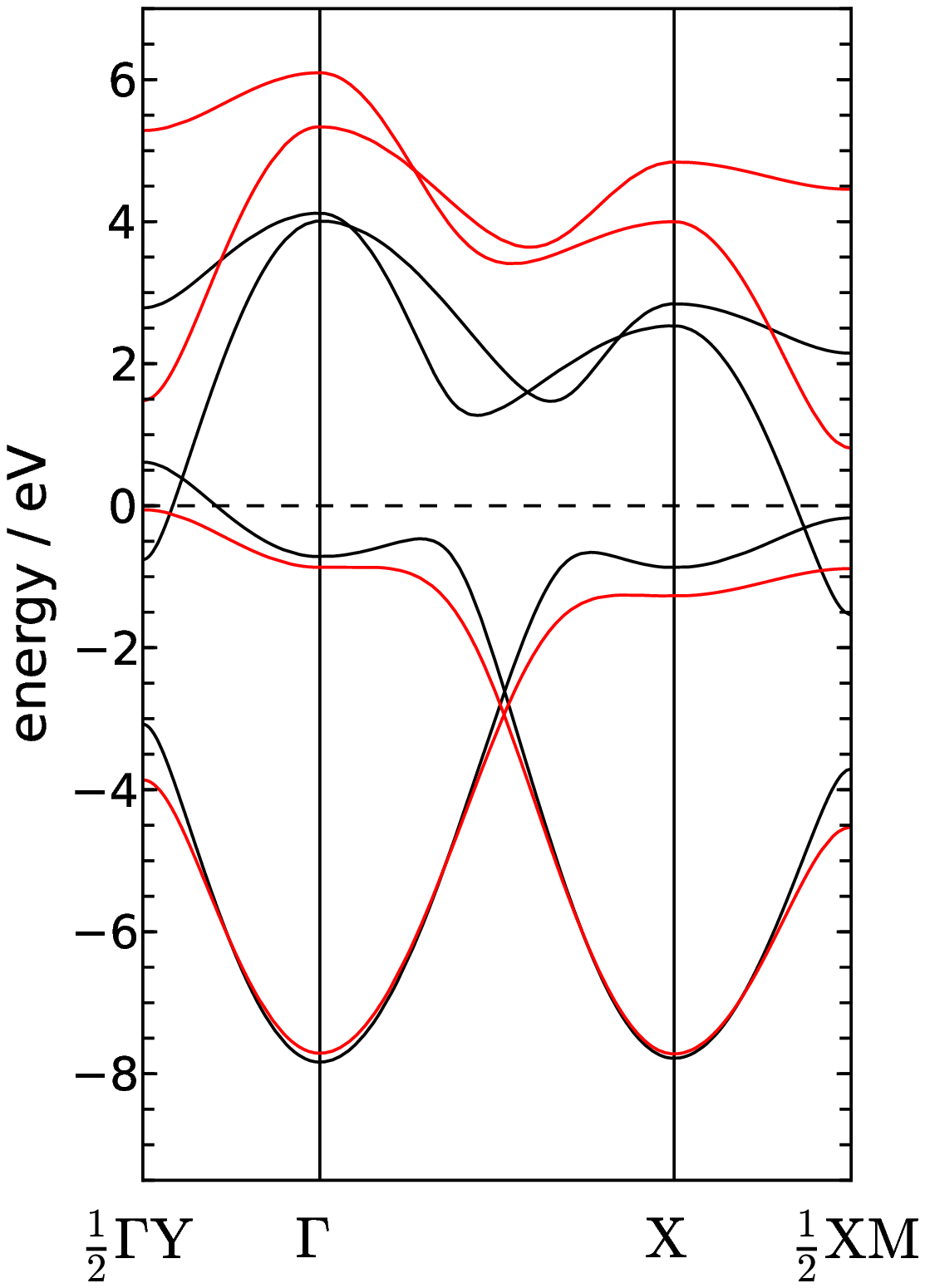} 

		\centering{(b)}
	\end{minipage}
	\caption{
		Bands in the folded Brillouin zone: (a) bare bands 
		with the Dirac points along the 
		1/2 $\Gamma\mathrm{Y}$-$\Gamma$ line and along the
		$\mathrm{X}$-1/2 $\mathrm{XM}$ line; (b) bands from Gutzwiller calculations
		with $U=7.0\,\mathrm{eV}$ and $J=1.0\,\mathrm{eV}$ (black) and with
		$U=9.0\,\mathrm{eV}$ and $J=1.0\,\mathrm{eV}$ (red). Note that the
		crossings along the $\Gamma$-$\mathrm{X}$ line are lifted due to
		the local interaction while the Dirac points persist until the
		$\mathrm{XZ}$ level is shifted above the $\mathrm{YZ}$ level at the
		corners of the BZ. This results in the zipper-like behavior in the
		opening of the band gap.}
	\label{fig:bandsGutzwiller} 
\end{figure}

The unperturbed band structure for the antiferromagnetic irreducible Brillouin
zone is shown in black in
Fig.~\ref{fig:bandsGutzwiller}\,$(a)$ and the Fermi 
level is chosen to be zero. The bands display four
crossings: one along the $1/2\Gamma\mathrm{Y}-\Gamma$ direction,
two along the $\Gamma-\mathrm{X}$ direction and one along the
$\mathrm{X}-1/2\mathrm{XM}$ direction. Comparing these bands with the 
bands from Fig.~\ref{fig:bandsGutzwiller}\,$(b)$ shows the influence of local interactions
onto the band structure. The black bands in Fig.~\ref{fig:bandsGutzwiller}\,$(b)$ result from
a Gutzwiller calculation with $U=7.0\,\mathrm{eV}=0.55\,W$ and $J=1.0\,\mathrm{eV}$. One observes that exchange gaps in
the electronic spectrum open only along the $\Gamma-\mathrm{X}$ direction. These
are the crossings where the orbital characters of the folded states at the Fermi
surface match.  The two remaining crossings result from bands with different
orbital character. As a consequence, no exchange gap opens along the
$1/2\Gamma\mathrm{Y}-\Gamma$ and the $\mathrm{X}-1/2\mathrm{XM}$ line, but the
folded bands form \textit{Dirac} cones in the dispersion. The topological
reasons for this behavior are explained in Ref.~\onlinecite{Ran:2009} However,
the results in Fig.~\ref{fig:bandsGutzwiller}\,$(b)$ show that the 
perturbation by local Coulomb interaction with intermediate size does not
destroy this property and the system stays metallic. 

%%%%%%
%% Figure 2: Fermi surface
%%%%%%

Increasing the local interaction
shifts the Dirac points towards the $1/2\Gamma\mathrm{Y}$ and the
$1/2\mathrm{XM}$ point, respectively. 
This zipper-like behavior of the opening of the gap persists until the $XZ$ level
at the corners of the SDW BZ is pushed above the $YZ$ one.  At this stage a gap
opens in the electronic spectrum and leads to a state that is an
antiferromagnetic Slater insulator (AFMSI).
This is shown by the red bands in Fig.~\ref{fig:bandsGutzwiller}\,$(b)$.
These bands result from a calculation with $U=9.0\,\mathrm{eV}=0.7\,W$ and
$J=1.0\,\mathrm{eV}$. Considering the opening of the gap, one observes a different
behavior for $J=0$ and for a finite $J$. For $J=0$ the gap opens
by a sudden jump to a large value of $8.4\,\mathrm{eV}$, while for finite $J$, 
\textit{e.g.}, $J=1.0\,\mathrm{eV}$, the gap opens continuously. The critical
value of the Hubbard interaction for this kind of phase transition will be 
denoted by $U_S$. This value is also a function of the Hund's 
rule coupling, \textit{i.e.},~$U_S=U_S(J)$.

\begin{figure}[htb]
	\includegraphics[width=0.41\textwidth]{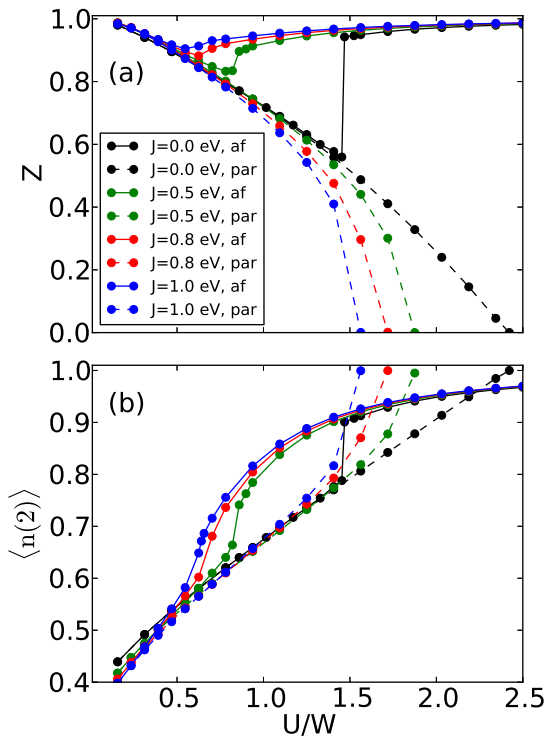}
	\caption{Paramagnetic (dashed
		lines) and antiferromagnetic (full lines) results from Gutzwiller
		calculations for the model for $J=0$ (black), $J=0.5\,\mathrm{eV}$ (green),
		$J=0.8\,\mathrm{eV}$ (red) and
		$J=1.0\,\mathrm{eV}$ (blue);
		(a) quasi-particle weight $Z$, (b) population of 
		the local charge state $n=2$.}
	\label{fig:gutzI}
\end{figure}

\begin{figure}[htb]
	\includegraphics[width=0.41\textwidth]{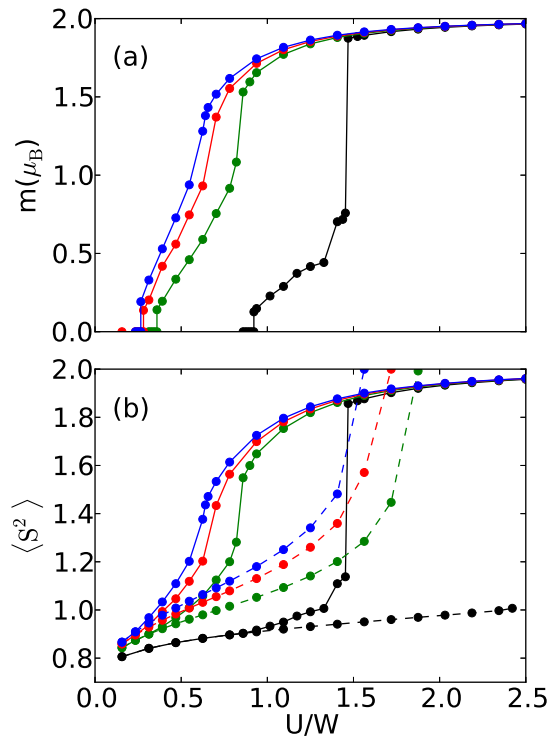}
	\caption{Gutzwiller calculation results for the model at $N=2$;
			(a) magnetization $m$, (b) average local moment $S^2$. 
			Denotation of lines as in Fig.~\ref{fig:gutzI}.
			}
	\label{fig:gutzII} 
\end{figure}

In order to compare further the paramagnetic and antiferromagnetic phases, I 
address the quasi-particle weight $Z$ and the average number of 
atomic configurations with two electrons $n(2)\equiv \langle n(2)\rangle$ in 
Fig.~\ref{fig:gutzI}. Both quantities serve as indicators of a Mott insulator
transition. Furthermore, in Fig.~\ref{fig:gutzII}, I give
results for the ordered moment $m$ and the local moment 
$\langle \mathbf{S}^2\rangle$. All results are plotted as
a function of $U$, for fixed values of $J=0$ (black), $0.5\,\mathrm{eV}$ (green),
$0.8\,\mathrm{eV}$ (red) and $1.0\,\mathrm{eV}$ (blue).  

%%%%%%
%% Figure 3: Gutzwiller results
%%%%%%

In Fig.~\ref{fig:gutzI}\,$(a)$ I show the quasi-particle renormalization $Z$, for both
the paramagnetic (dashed) and the antiferromagnetic solutions. As already seen in 
Fig.~\ref{fig:gwpara}, in the paramagnetic case, I find a Brinkman-Rice transition 
at some critical value $U_c$. This critical value depends on the Hund's rule
exchange and for $N=2$ a larger Hund's rule exchange decreases $U_c$. 
The situation changes drastically in the presence of antiferromagnetic stripe
order. Considering the quasi-particle renormalization $Z$, the antiferromagnetic
solution closely follows the paramagnetic one for not too large $U$ ($U \leq
0.5\,W$). As the long-range 
magnetic order sets in, the values for $Z$ become larger than in the
paramagnetic case, \textit{i.e.}, correlations become weaker. 
Increasing $U$ further leads to a jump in $Z$ towards unity.
The reason for such a behavior is the metal insulator
transition of Slater type at $U_S$ which was already discussed. 
One observes that for larger
values of the Hund's rule coupling the MIT occurs for smaller values
of $U$. As a consequence, the change in the quasi-particle renormalization
is less dramatic for larger values of $J$.

%%%%%%
%% Figure 4: Gutzwiller results
%%%%%%

In Fig.~\ref{fig:gutzI}\,($b$), I plot the expectation value of the atomic configurations with two
electrons ($n(2)$). In the paramagnetic
case, $n(2)$ starts from the non-interacting limit $n(2)=3/8$ and saturates to
unity at the MIT, where all charge fluctuations are frozen (Brinkman-Rice
localization transition). A comparison of Fig.~\ref{fig:gutzI}\,($a$) and
Fig.~\ref{fig:gutzI}\,($b$) shows that the behavior of $n(2)$ follows closely that of 
the quasi-particle weight $Z$. As in the case of the quasi-particle renormalization
$Z$, the behavior of $n(2)$ is strongly influenced by the 
antiferromagnetic order. Here, $n(2)$ starts by following the paramagnetic curve,
but it departs around the MIT. While in the paramagnetic case with a
Mott-Hubbard MIT $n(2)$ goes to unity, in the antiferromagnetic case $n(2)$ only
increases slowly.  It saturates to unity at values of $U$ much larger than the
critical $U_c$ for the paramagnetic MIT.  This again shows that the (Slater)
band-insulating regime is less correlated than the corresponding paramagnetic
regime.

In Fig.~\ref{fig:gutzII}\,($a$) I plot the long-range ordered magnetic moment $m$ as a function of
$U$ for different values of the Hund's rule coupling. The plot shows indeed that the point
where the paramagnetic and antiferromagnetic curve of $Z$ ($n(2)$) run apart, is the point where
long-range antiferromagnetic order sets in. Moreover, for every curve in Fig.~\ref{fig:gutzII}\,($a$)
one can observe a value of the Hubbard interaction with a sudden change of the slope of the $m$~vs~$U$-curve,
\textit{e.g.},~at $U \approx 1.5\,W$ for $J=0$. These points indicate the Slater MIT in the antiferromagnetic case.

In order to conclude these observations, in the parameter range of this 
model study only three phases are stable: a paramagnetic metal
for small values of $U,J$, an AFM metal for intermediate values and,
finally, an AFM band (Slater) insulator for $U \gg W$. This is 
in stark contrast to the phase diagram obtained by
the slave rotor approach for the same model in Ref.~\onlinecite{Ko:2011}. The phase diagram obtained by
this method is much richer compared to what is observed with the
variationally controlled Gutzwiller approach.

%%%%%%
%% Figure 5: Rescaling
%%%%%%
\begin{figure}[htbp]
	\includegraphics[width=0.41\textwidth]{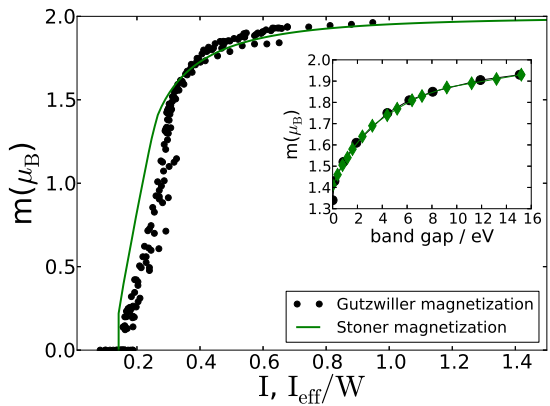}
	\caption{\label{fig:rescale}
		Ordered (black circles) magnetic moment in the AFM phase,
		as a function of the effective magnetic scale $I_{\mathrm{eff}}=J+U/3$ ($J >
		0.2\,\mathrm{eV}$); the $m$~vs~$I$ response of the Stoner model is shown
		as a line.
		Inset: magnetic moment as a function of the band gap. I compare results
		from the Stoner calculation (green) with results from Gutzwiller calculation
		(black) for $J=1.0\,\mathrm{eV}$. }
\end{figure}

Next, I discuss the relationship between the ordered magnetic moment
and the local magnetic moment in more detail. Fig.~\ref{fig:gutzII}\,($b$) shows the
average value of the local spin operator $\langle S^2 \rangle$, where the
dashed lines show the results in the paramagnetic case.  For the local
moment one has to distinguish the cases $J=0$ and finite values of $J$.
For $J=0$, the value $\langle S^2 \rangle$ as a function of $U$ increases from the
non-interacting limit $\langle S^2 \rangle = 3/4$ only slowly. In contrast
to this, for finite $J$, one observes a strong increase to the fully localized
value  $\langle S^2 \rangle = 2$ at the Mott-Hubbard MIT.  Again, the
results change substantially when one allows for antiferromagnetic order. Only for
$J=0$ there is a sudden increase of $\langle S^2 \rangle$ at the Slater MIT.
For a finite value of $J$, however, the dependence of
$\langle S^2 \rangle$ on $U$ is smooth. It is interesting to note that, for $U
\gtrsim 15\,\mathrm{eV} \sim 1.2\,W$ and for finite $J$, the local magnetic moment is larger in the paramagnetic than in the
antiferromagnetic phase.  This seemingly counter-intuitive result is also found
in a recent DMFT study of the single band Hubbard model.~\cite{Taranto:2012,
Toschi:2012} In this publication, it it also reported that in the 
weak-coupling regime the antiferromagnetic order is stabilized by a
gain in potential energy, while in the strong-coupling regime a gain in kinetic energy
leads to the stabilization of the magnetic phase. The results of the current work agree
with this observation.

Finally, I consider the shape of the curves $m(U)$ in
Fig.~\ref{fig:gutzII}\,($a$) and the shape of $\langle S^2 \rangle$ as
a function of $U$~(Fig.~\ref{fig:gutzII}\,($b$)) in the antiferromagnetic case for finite $J$. All curves strongly
resemble each other and, as seen in Fig.~\ref{fig:gutzI}\,($a$), the quasi-particle renormalization is close
to unity. This indicates that a band picture for quasi-particles
will provide a reasonable description. In order to illustrate this 
concept, I show the results of a Stoner mean-field calculation for
the half-filled two-band model in Fig.~\ref{fig:rescale}. Here, the magnetization
$m$ is shown as a function of the Stoner parameter $I$ (green line).
The two quantities are related to
each other via the self-consistency condition
$\Delta = m I$ with the splitting parameter $\Delta$.
The plot shows that the $m(I)$ curve has a continuous and
monotonous behavior between $m=0$ and $m=2\,\mu_{\mathrm{B}}$.  The inflection point 
at $m \sim 1.41\,\mu_{\mathrm{B}}$ coincides with the metal-insulator
transition, \textit{i.e.}, with the point, where the exchange gap is large enough to
eliminate the Dirac points.  It corresponds to $I=3.43\,\mathrm{eV}$
and $\Delta=4.8\,\mathrm{eV}$.  

In Fig.~\ref{fig:rescale} I also plot the ordered magnetic moment 
from the Gutzwiller calculations as a function of the effective energy scale $I_{\mathrm{eff}}=J+U/3$.
The data for small $J$ are anomalous, therefore I just consider results
with $J > 0.2\,\textrm{eV}$. By choosing the energy scale $I_{\mathrm{eff}}$, 
all the curves $m(U,J)$ fall on top of each other.
The inset of Fig.~\ref{fig:rescale} shows the magnetization as a function of the
band gap. 
The results are from the Stoner calculation (green) and from the
Gutzwiller calculation with $J=1.0\,\mathrm{eV}$ (black). 
In principle, the dependence of the ordered magnetic moment on the band width could
be tested experimentally. As these two curves fall
onto each other, also from this point of view there is no fundamental difference between
the Stoner results and the Gutzwiller results.
From these observations one can draw three important
conclusions:
\begin{enumerate}
	\item Except for small $J$, which is anomalous,
	there is a unique, \textit{atomic} scale governing magnetism, both 
	for the ordered and local moment. In the present case
	it is $I_{\mathrm{eff}}=J+U/3$.
	\item The value of the magnetic moment, both ordered and local, 
	is set by the band response of the system. 
	\item This effect is independent on whether the AFM solution 
	is stabilized by kinetic or potential energy, Ref.~\onlinecite{Taranto:2012}.
\end{enumerate}
Note in particular, that $U$ is considerably larger
than $J$. Therefore, the effective energy scale 
$I_{\mathrm{eff}}$ is mainly determined
by the value of the Hubbard interaction. This is in strong contrast to the recent Gutzwiller study 
on an eight-band model for LaOFeAs. In that study 
the effective energy scale for the magnetism which was mainly
determined by the Hund's coupling $J$.~\cite{Schickling:2012}
I consider this as an important point as it discloses
\textsl{qualitative} differences between the minimal two-band model 
and more realistic models.

To summarize, I studied a two-band model for the iron
pnictides. This model contains two important properties of this
class of materials: the strong nesting that leads to a stripe-ordered
spin-density wave and the
Dirac points. I showed how antiferromagnetism evolves in
this model due to correlations and how the gain of energy by the
antiferromagnetic order reduces the effects of the local 
correlations. The magnetism itself is determined 
by an effective energy scale $I_{\mathrm{eff}}=J+U/3$. Since
$I_{\mathrm{eff}}$ is large, of the order of $U$, all results
reflect Stoner-type band magnetism.  Moreover, as
$I_{\mathrm{eff}}$ depends marginally on the Hund's coupling the
model does not describe a Hund's metal. This observation is contrast
to studies on more realistic models.
For large interactions the model describes an antiferromagnetic
band insulator. The influence of the Dirac points is important
as they delay the opening of the band gap.

\acknowledgments
I thank L.~Boeri, F.~Gebhard, J.~B\"unemann, L. de' Medici, A.~Toschi and C.~Taranto for useful 
comments and discussion. I also acknowledge hospitality of the Max Planck
Institute for Solid State Research where parts of this work were performed.

\bibliography{twoBand}

%Merlin.mbs v4.21 2009-07-09.
\begin{thebibliography}{10}%
\makeatletter
\providecommand \@ifxundefined [1]{%
 \ifx #1\undefined \expandafter \@firstoftwo
 \else \expandafter \@secondoftwo
\fi
}%
\providecommand \@ifnum [1]{%
 \ifnum #1\expandafter \@firstoftwo
 \else \expandafter \@secondoftwo
\fi
}%
\providecommand \enquote [1]{``#1''}%
\providecommand \bibnamefont  [1]{#1}%
\providecommand \bibfnamefont [1]{#1}%
\providecommand \citenamefont [1]{#1}%
\providecommand\href[0]{\@sanitize\@href}%
\providecommand\@href[1]{\endgroup\@@startlink{#1}\endgroup\@@href}%
\providecommand\@@href[1]{#1\@@endlink}%
\providecommand \@sanitize [0]{\begingroup\catcode`\&12\catcode`\#12\relax}%
\@ifxundefined \pdfoutput {\@firstoftwo}{%
 \@ifnum{\z@=\pdfoutput}{\@firstoftwo}{\@secondoftwo}%
}{%
 \providecommand\@@startlink[1]{\leavevmode\special{html:<a href="#1">}}%
 \providecommand\@@endlink[0]{\special{html:</a>}}%
}{%
 \providecommand\@@startlink[1]{%
  \leavevmode
  \pdfstartlink
   attr{/Border[0 0 1 ]/H/I/C[0 1 1]}%
   user{/Subtype/Link/A<</Type/Action/S/URI/URI(#1)>>}%
  \relax
 }%
 \providecommand\@@endlink[0]{\pdfendlink}%
}%
\providecommand \url  [0]{\begingroup\@sanitize \@url }%
\providecommand \@url [1]{\endgroup\@href {#1}{\urlprefix}}%
\providecommand \urlprefix [0]{URL }%
\providecommand \Eprint[0]{\href }%
\@ifxundefined \urlstyle {%
  \providecommand \doi [1]{doi:\discretionary{}{}{}#1}%
}{%
  \providecommand \doi [0]{doi:\discretionary{}{}{}\begingroup
  \urlstyle{rm}\Url }%
}%
\providecommand \doibase [0]{http://dx.doi.org/}%
\providecommand \Doi[1]{\href{\doibase#1}}%
\providecommand \bibAnnote [3]{%
  \BibitemShut{#1}%
  \begin{quotation}\noindent
    \textsc{Key:}\ #2\\\textsc{Annotation:}\ #3%
  \end{quotation}%
}%
\providecommand \bibAnnoteFile [2]{%
  \IfFileExists{#2}{\bibAnnote {#1} {#2} {\input{#2}}}{}%
}%
\providecommand \typeout [0]{\immediate \write \m@ne }%
\providecommand \selectlanguage [0]{\@gobble}%
\providecommand \bibinfo [0]{\@secondoftwo}%
\providecommand \bibfield [0]{\@secondoftwo}%
\providecommand \translation [1]{[#1]}%
\providecommand \BibitemOpen[0]{}%
\providecommand \bibitemStop [0]{}%
\providecommand \bibitemNoStop [0]{.\EOS\space}%
\providecommand \EOS [0]{\spacefactor3000\relax}%
\providecommand \BibitemShut [1]{\csname bibitem#1\endcsname}%
%</preamble>
\bibitem{Mazin:2008a}%
  \BibitemOpen
  \bibfield{author}{%
  \bibinfo {author} {\bibfnamefont{I.~I.}\ \bibnamefont{Mazin}}, \bibinfo
  {author} {\bibfnamefont{D.~J.}\ \bibnamefont{Singh}}, \bibinfo {author}
  {\bibfnamefont{M.~D.}\ \bibnamefont{Johannes}},\ and\ \bibinfo {author}
  {\bibfnamefont{M.-H.}\ \bibnamefont{Du}},\ }%
  \bibfield{journal}{%
  \Doi{10.1103/PhysRevLett.101.057003}{\bibinfo {journal} {Phys. Rev. Lett.}}\
  }%
  \textbf{\bibinfo {volume} {101}},\ \bibinfo {pages} {057003} (\bibinfo {year}
  {2008})%
  \bibAnnoteFile{NoStop}{Mazin:2008a}%
\bibitem{Singh:2008}%
  \BibitemOpen
  \bibfield{author}{%
  \bibinfo {author} {\bibfnamefont{D.~J.}\ \bibnamefont{Singh}}\ and\ \bibinfo
  {author} {\bibfnamefont{M.-H.}\ \bibnamefont{Du}},\ }%
  \bibfield{journal}{%
  \Doi{10.1103/PhysRevLett.100.237003}{\bibinfo {journal} {Phys. Rev. Lett.}}\
  }%
  \textbf{\bibinfo {volume} {100}},\ \bibinfo {pages} {237003} (\bibinfo {year}
  {2008})%
  \bibAnnoteFile{NoStop}{Singh:2008}%
\bibitem{Yildirim:2008}%
  \BibitemOpen
  \bibfield{author}{%
  \bibinfo {author} {\bibfnamefont{T.}~\bibnamefont{Yildirim}},\ }%
  \bibfield{journal}{%
  \Doi{10.1103/PhysRevLett.101.057010}{\bibinfo {journal} {Phys. Rev. Lett.}}\
  }%
  \textbf{\bibinfo {volume} {101}},\ \bibinfo {pages} {057010} (\bibinfo {year}
  {2008})%
  \bibAnnoteFile{NoStop}{Yildirim:2008}%
\bibitem{Ishida:2010}%
  \BibitemOpen
  \bibfield{author}{%
  \bibinfo {author} {\bibfnamefont{H.}~\bibnamefont{Ishida}}\ and\ \bibinfo
  {author} {\bibfnamefont{A.}~\bibnamefont{Liebsch}},\ }%
  \bibfield{journal}{%
  \Doi{10.1103/PhysRevB.81.054513}{\bibinfo {journal} {Phys. Rev. B}}\ }%
  \textbf{\bibinfo {volume} {81}},\ \bibinfo {pages} {054513} (\bibinfo {year}
  {2010})%
  \bibAnnoteFile{NoStop}{Ishida:2010}%
\bibitem{Aichhorn:2011}%
  \BibitemOpen
  \bibfield{author}{%
  \bibinfo {author} {\bibfnamefont{M.}~\bibnamefont{Aichhorn}}, \bibinfo
  {author} {\bibfnamefont{L.}~\bibnamefont{Pourovskii}},\ and\ \bibinfo
  {author} {\bibfnamefont{A.}~\bibnamefont{Georges}},\ }%
  \bibfield{journal}{%
  \Doi{10.1103/PhysRevB.84.054529}{\bibinfo {journal} {Phys. Rev. B}}\ }%
  \textbf{\bibinfo {volume} {84}},\ \bibinfo {pages} {054529} (\bibinfo {year}
  {2011})%
  \bibAnnoteFile{NoStop}{Aichhorn:2011}%
\bibitem{Yin:2011_b}%
  \BibitemOpen
  \bibfield{author}{%
  \bibinfo {author} {\bibfnamefont{Z.~P.}\ \bibnamefont{Yin}}, \bibinfo
  {author} {\bibfnamefont{K.}~\bibnamefont{Haule}},\ and\ \bibinfo {author}
  {\bibfnamefont{G.}~\bibnamefont{Kotliar}},\ }%
  \bibfield{journal}{%
  \Doi{10.1038/nmat3120}{\bibinfo {journal} {Nature materials}}\ }%
  \textbf{\bibinfo {volume} {10}},\ \bibinfo {pages} {1} (\bibinfo {year}
  {2011})%
  \bibAnnoteFile{NoStop}{Yin:2011_b}%
\bibitem{Yao:2011_a}%
  \BibitemOpen
  \bibfield{author}{%
  \bibinfo {author} {\bibfnamefont{Y.~X.}\ \bibnamefont{Yao}}, \bibinfo
  {author} {\bibfnamefont{J.}~\bibnamefont{Schmalian}}, \bibinfo {author}
  {\bibfnamefont{C.~Z.}\ \bibnamefont{Wang}}, \bibinfo {author}
  {\bibfnamefont{K.~M.}\ \bibnamefont{Ho}},\ and\ \bibinfo {author}
  {\bibfnamefont{G.}~\bibnamefont{Kotliar}},\ }%
  \bibfield{journal}{%
  \Doi{10.1103/PhysRevB.84.245112}{\bibinfo {journal} {Phys. Rev. B}}\ }%
  \textbf{\bibinfo {volume} {84}},\ \bibinfo {pages} {245112} (\bibinfo {year}
  {2011})%
  \bibAnnoteFile{NoStop}{Yao:2011_a}%
\bibitem{Schickling:2011}%
  \BibitemOpen
  \bibfield{author}{%
  \bibinfo {author} {\bibfnamefont{T.}~\bibnamefont{Schickling}}, \bibinfo
  {author} {\bibfnamefont{F.}~\bibnamefont{Gebhard}},\ and\ \bibinfo {author}
  {\bibfnamefont{J.}~\bibnamefont{B\"{u}nemann}},\ }%
  \bibfield{journal}{%
  \Doi{10.1103/PhysRevLett.106.146402}{\bibinfo {journal} {Phys. Rev. Lett.}}\
  }%
  \textbf{\bibinfo {volume} {106}},\ \bibinfo {pages} {146402} (\bibinfo {year}
  {2011})%
  \bibAnnoteFile{NoStop}{Schickling:2011}%
\bibitem{Schickling:2012}%
  \BibitemOpen
  \bibfield{author}{%
  \bibinfo {author} {\bibfnamefont{T.}~\bibnamefont{Schickling}}, \bibinfo
  {author} {\bibfnamefont{F.}~\bibnamefont{Gebhard}}, \bibinfo {author}
  {\bibfnamefont{J.}~\bibnamefont{B\"{u}nemann}}, \bibinfo {author}
  {\bibfnamefont{L.}~\bibnamefont{Boeri}}, \bibinfo {author}
  {\bibfnamefont{O.~K.}\ \bibnamefont{Andersen}},\ and\ \bibinfo {author}
  {\bibfnamefont{W.}~\bibnamefont{Weber}},\ }%
  \bibfield{journal}{%
  \Doi{10.1103/PhysRevLett.108.036406}{\bibinfo {journal} {Phys. Rev. Lett.}}\
  }%
  \textbf{\bibinfo {volume} {108}},\ \bibinfo {pages} {036406} (\bibinfo {year}
  {2012})%
  \bibAnnoteFile{NoStop}{Schickling:2012}%
\bibitem{Mazin:2008c}%
  \BibitemOpen
  \bibfield{author}{%
  \bibinfo {author} {\bibfnamefont{I.~I.}\ \bibnamefont{Mazin}}, \bibinfo
  {author} {\bibfnamefont{M.~D.}\ \bibnamefont{Johannes}}, \bibinfo {author}
  {\bibfnamefont{L.}~\bibnamefont{Boeri}}, \bibinfo {author}
  {\bibfnamefont{K.}~\bibnamefont{Koepernik}},\ and\ \bibinfo {author}
  {\bibfnamefont{D.~J.}\ \bibnamefont{Singh}},\ }%
  \bibfield{journal}{%
  \Doi{10.1103/PhysRevB.78.085104}{\bibinfo {journal} {Phys. Rev. B}}\ }%
  \textbf{\bibinfo {volume} {78}},\ \bibinfo {pages} {085104} (\bibinfo {year}
  {2008})%
  \bibAnnoteFile{NoStop}{Mazin:2008c}%
\bibitem{Si:2008}%
  \BibitemOpen
  \bibfield{author}{%
  \bibinfo {author} {\bibfnamefont{Q.}~\bibnamefont{Si}}\ and\ \bibinfo
  {author} {\bibfnamefont{E.}~\bibnamefont{Abrahams}},\ }%
  \bibfield{journal}{%
  \Doi{10.1103/PhysRevLett.101.076401}{\bibinfo {journal} {Phys. Rev. Lett.}}\
  }%
  \textbf{\bibinfo {volume} {101}},\ \bibinfo {pages} {076401} (\bibinfo {year}
  {2008})%
  \bibAnnoteFile{NoStop}{Si:2008}%
\bibitem{Stanek:2011}%
  \BibitemOpen
  \bibfield{author}{%
  \bibinfo {author} {\bibfnamefont{D.}~\bibnamefont{Stanek}}, \bibinfo {author}
  {\bibfnamefont{O.~P.}\ \bibnamefont{Sushkov}},\ and\ \bibinfo {author}
  {\bibfnamefont{G.~S.}\ \bibnamefont{Uhrig}},\ }%
  \bibfield{journal}{%
  \Doi{10.1103/PhysRevB.84.064505}{\bibinfo {journal} {Phys. Rev. B}}\ }%
  \textbf{\bibinfo {volume} {84}},\ \bibinfo {pages} {064505} (\bibinfo {year}
  {2011})%
  \bibAnnoteFile{NoStop}{Stanek:2011}%
\bibitem{Hu:2012}%
  \BibitemOpen
  \bibfield{author}{%
  \bibinfo {author} {\bibfnamefont{J.}~\bibnamefont{Hu}}\ and\ \bibinfo
  {author} {\bibfnamefont{N.}~\bibnamefont{Hao}},\ }%
  \bibfield{journal}{%
  \bibinfo {journal} {Phys. Rev. X}\ }%
  \textbf{\bibinfo {volume} {2}},\ \bibinfo {pages} {21009} (\bibinfo {year}
  {2012})%
  \bibAnnoteFile{NoStop}{Hu:2012}%
\bibitem{Lo:2013}%
  \BibitemOpen
  \bibfield{author}{%
  \bibinfo {author} {\bibfnamefont{K.~W.}\ \bibnamefont{Lo}}, \bibinfo {author}
  {\bibfnamefont{W.-C.}\ \bibnamefont{Lee}},\ and\ \bibinfo {author}
  {\bibfnamefont{P.~W.}\ \bibnamefont{Phillips}},\ }%
  \bibfield{journal}{%
  \bibinfo {journal} {EPL}\ }%
  \textbf{\bibinfo {volume} {101}},\ \bibinfo {pages} {50007} (\bibinfo {year}
  {2013})%
  \bibAnnoteFile{NoStop}{Lo:2013}%
\bibitem{Tai:2013}%
  \BibitemOpen
  \bibfield{author}{%
  \bibinfo {author} {\bibfnamefont{Y.-Y.}\ \bibnamefont{Tai}}, \bibinfo
  {author} {\bibfnamefont{J.-X.}\ \bibnamefont{Zhu}}, \bibinfo {author}
  {\bibfnamefont{M.~J.}\ \bibnamefont{Graf}},\ and\ \bibinfo {author}
  {\bibfnamefont{C.~S.}\ \bibnamefont{Ting}},\ }%
  \bibfield{journal}{%
  \bibinfo {journal} {EPL}\ }%
  \textbf{\bibinfo {volume} {103}},\ \bibinfo {pages} {67001} (\bibinfo {year}
  {2013})%
  \bibAnnoteFile{NoStop}{Tai:2013}%
\bibitem{Quan:2012}%
  \BibitemOpen
  \bibfield{author}{%
  \bibinfo {author} {\bibfnamefont{Y.-M.}\ \bibnamefont{Quan}}, \bibinfo
  {author} {\bibfnamefont{L.-J.}\ \bibnamefont{Zou}}, \bibinfo {author}
  {\bibfnamefont{D.-Y.}\ \bibnamefont{Liu}},\ and\ \bibinfo {author}
  {\bibfnamefont{H.-Q.}\ \bibnamefont{Lin}},\ }%
  \bibfield{journal}{%
  \bibinfo {journal} {Journal of Physics: Condensed Matter}\ }%
  \textbf{\bibinfo {volume} {24}},\ \bibinfo {pages} {85603} (\bibinfo {year}
  {2012})%
  \bibAnnoteFile{NoStop}{Quan:2012}%
\bibitem{Ran:2009}%
  \BibitemOpen
  \bibfield{author}{%
  \bibinfo {author} {\bibfnamefont{Y.}~\bibnamefont{Ran}}, \bibinfo {author}
  {\bibfnamefont{F.}~\bibnamefont{Wang}}, \bibinfo {author}
  {\bibfnamefont{H.}~\bibnamefont{Zhai}}, \bibinfo {author}
  {\bibfnamefont{A.}~\bibnamefont{Vishwanath}},\ and\ \bibinfo {author}
  {\bibfnamefont{D.-H.}\ \bibnamefont{Lee}},\ }%
  \bibfield{journal}{%
  \Doi{10.1103/PhysRevB.79.014505}{\bibinfo {journal} {Phys. Rev. B}}\ }%
  \textbf{\bibinfo {volume} {79}},\ \bibinfo {pages} {014505} (\bibinfo {year}
  {2009})%
  \bibAnnoteFile{NoStop}{Ran:2009}%
\bibitem{Ko:2011}%
  \BibitemOpen
  \bibfield{author}{%
  \bibinfo {author} {\bibfnamefont{W.-H.}\ \bibnamefont{Ko}}\ and\ \bibinfo
  {author} {\bibfnamefont{P.~A.}\ \bibnamefont{Lee}},\ }%
  \bibfield{journal}{%
  \Doi{10.1103/PhysRevB.83.134515}{\bibinfo {journal} {Phys. Rev. B}}\ }%
  \textbf{\bibinfo {volume} {83}},\ \bibinfo {pages} {134515} (\bibinfo {year}
  {2011})%
  \bibAnnoteFile{NoStop}{Ko:2011}%
\bibitem{Buenemann_2012_b}%
  \BibitemOpen
  \bibfield{author}{%
  \bibinfo {author} {\bibfnamefont{J.}~\bibnamefont{B\"{u}nemann}}, \bibinfo
  {author} {\bibfnamefont{F.}~\bibnamefont{Gebhard}}, \bibinfo {author}
  {\bibfnamefont{T.}~\bibnamefont{Schickling}},\ and\ \bibinfo {author}
  {\bibfnamefont{W.}~\bibnamefont{Weber}},\ }%
  \bibfield{journal}{%
  \Doi{10.1002/pssb.201147585}{\bibinfo {journal} {physica status solidi (b)}}\
  }%
  \textbf{\bibinfo {volume} {249}},\ \bibinfo {pages} {1282} (\bibinfo {year}
  {2012}),\ ISSN \bibinfo {issn} {1521-3951}%
  \bibAnnoteFile{NoStop}{Buenemann_2012_b}%
\bibitem{Buenemann:1998}%
  \BibitemOpen
  \bibfield{author}{%
  \bibinfo {author} {\bibfnamefont{J.}~\bibnamefont{B\"{u}nemann}}, \bibinfo
  {author} {\bibfnamefont{W.}~\bibnamefont{Weber}},\ and\ \bibinfo {author}
  {\bibfnamefont{F.}~\bibnamefont{Gebhard}},\ }%
  \bibfield{journal}{%
  \Doi{10.1103/PhysRevB.57.6896}{\bibinfo {journal} {Phys. Rev. B}}\ }%
  \textbf{\bibinfo {volume} {57}},\ \bibinfo {pages} {6896} (\bibinfo {year}
  {1998})%
  \bibAnnoteFile{NoStop}{Buenemann:1998}%
\bibitem{Buenemann:2005}%
  \BibitemOpen
  \bibfield{author}{%
  \bibinfo {author} {\bibfnamefont{J.}~\bibnamefont{B\"unemann}}, \bibinfo
  {author} {\bibfnamefont{F.}~\bibnamefont{Gebhard}}, \bibinfo {author}
  {\bibfnamefont{T.}~\bibnamefont{Ohm}}, \bibinfo {author}
  {\bibfnamefont{S.}~\bibnamefont{Weiser}},\ and\ \bibinfo {author}
  {\bibfnamefont{W.}~\bibnamefont{Weber}},\ }%
  in\ \emph{\bibinfo {booktitle} {Frontiers in Magnetic Materials}},\ \bibinfo
  {editor} {edited by\ \bibinfo {editor} {\bibfnamefont{A.~V.}\
  \bibnamefont{Narlikar}}}\ (\bibinfo {publisher} {Springer, Berlin},\ \bibinfo
  {year} {2005})\ pp.\ \bibinfo {pages} {117--151}%
  \bibAnnoteFile{NoStop}{Buenemann:2005}%
\bibitem{Buenemann:2003}%
  \BibitemOpen
  \bibfield{author}{%
  \bibinfo {author} {\bibfnamefont{J.}~\bibnamefont{B\"unemann}}, \bibinfo
  {author} {\bibfnamefont{F.}~\bibnamefont{Gebhard}},\ and\ \bibinfo {author}
  {\bibfnamefont{R.}~\bibnamefont{Thul}},\ }%
  \bibfield{journal}{%
  \Doi{10.1103/PhysRevB.67.075103}{\bibinfo {journal} {Phys. Rev. B}}\ }%
  \textbf{\bibinfo {volume} {67}},\ \bibinfo {pages} {075103} (\bibinfo {year}
  {2003})%
  \bibAnnoteFile{NoStop}{Buenemann:2003}%
\bibitem{Hofmann:2009}%
  \BibitemOpen
  \bibfield{author}{%
  \bibinfo {author} {\bibfnamefont{A.}~\bibnamefont{Hofmann}}, \bibinfo
  {author} {\bibfnamefont{X.~Y.}\ \bibnamefont{Cui}}, \bibinfo {author}
  {\bibfnamefont{J.}~\bibnamefont{Sch\"afer}}, \bibinfo {author}
  {\bibfnamefont{S.}~\bibnamefont{Meyer}}, \bibinfo {author}
  {\bibfnamefont{P.}~\bibnamefont{H\"opfner}}, \bibinfo {author}
  {\bibfnamefont{C.}~\bibnamefont{Blumenstein}}, \bibinfo {author}
  {\bibfnamefont{M.}~\bibnamefont{Paul}}, \bibinfo {author}
  {\bibfnamefont{L.}~\bibnamefont{Patthey}}, \bibinfo {author}
  {\bibfnamefont{E.}~\bibnamefont{Rotenberg}}, \bibinfo {author}
  {\bibfnamefont{J.}~\bibnamefont{B\"unemann}}, \bibinfo {author}
  {\bibfnamefont{F.}~\bibnamefont{Gebhard}}, \bibinfo {author}
  {\bibfnamefont{T.}~\bibnamefont{Ohm}}, \bibinfo {author}
  {\bibfnamefont{W.}~\bibnamefont{Weber}},\ and\ \bibinfo {author}
  {\bibfnamefont{R.}~\bibnamefont{Claessen}},\ }%
  \bibfield{journal}{%
  \Doi{10.1103/PhysRevLett.102.187204}{\bibinfo {journal} {Phys. Rev. Lett.}}\
  }%
  \textbf{\bibinfo {volume} {102}},\ \bibinfo {pages} {187204} (\bibinfo {year}
  {2009})%
  \bibAnnoteFile{NoStop}{Hofmann:2009}%
\bibitem{Brinkman:1970}%
  \BibitemOpen
  \bibfield{author}{%
  \bibinfo {author} {\bibfnamefont{W.~F.}\ \bibnamefont{Brinkman}}\ and\
  \bibinfo {author} {\bibfnamefont{T.~M.}\ \bibnamefont{Rice}},\ }%
  \bibfield{journal}{%
  \Doi{10.1103/PhysRevB.2.4302}{\bibinfo {journal} {Phys. Rev. B}}\ }%
  \textbf{\bibinfo {volume} {2}},\ \bibinfo {pages} {4302} (\bibinfo {year}
  {1970})%
  \bibAnnoteFile{NoStop}{Brinkman:1970}%
\bibitem{deMedici:2011_a}%
  \BibitemOpen
  \bibfield{author}{%
  \bibinfo {author} {\bibfnamefont{L.}~\bibnamefont{de' Medici}}, \bibinfo
  {author} {\bibfnamefont{J.}~\bibnamefont{Mravlje}},\ and\ \bibinfo {author}
  {\bibfnamefont{A.}~\bibnamefont{Georges}},\ }%
  \bibfield{journal}{%
  \Doi{10.1103/PhysRevLett.107.256401}{\bibinfo {journal} {Phys. Rev. Lett.}}\
  }%
  \textbf{\bibinfo {volume} {107}},\ \bibinfo {pages} {256401} (\bibinfo {year}
  {2011})%
  \bibAnnoteFile{NoStop}{deMedici:2011_a}%
\bibitem{deMedici:2011_b}%
  \BibitemOpen
  \bibfield{author}{%
  \bibinfo {author} {\bibfnamefont{L.}~\bibnamefont{de' Medici}},\ }%
  \bibfield{journal}{%
  \Doi{10.1103/PhysRevB.83.205112}{\bibinfo {journal} {Phys. Rev. B}}\ }%
  \textbf{\bibinfo {volume} {83}},\ \bibinfo {pages} {205112} (\bibinfo {year}
  {2011})%
  \bibAnnoteFile{NoStop}{deMedici:2011_b}%
\bibitem{Andersen2011}%
  \BibitemOpen
  \bibfield{author}{%
  \bibinfo {author} {\bibfnamefont{O.~K.}\ \bibnamefont{Andersen}}\ and\
  \bibinfo {author} {\bibfnamefont{L.}~\bibnamefont{Boeri}},\ }%
  \bibfield{journal}{%
  \bibinfo {journal} {Annalen der Physik}\ }%
  \textbf{\bibinfo {volume} {523}},\ \bibinfo {pages} {8} (\bibinfo {year}
  {2011})%
  \bibAnnoteFile{NoStop}{Andersen2011}%
\bibitem{Taranto:2012}%
  \BibitemOpen
  \bibfield{author}{%
  \bibinfo {author} {\bibfnamefont{C.}~\bibnamefont{Taranto}}, \bibinfo
  {author} {\bibfnamefont{G.}~\bibnamefont{Sangiovanni}}, \bibinfo {author}
  {\bibfnamefont{K.}~\bibnamefont{Held}}, \bibinfo {author}
  {\bibfnamefont{M.}~\bibnamefont{Capone}}, \bibinfo {author}
  {\bibfnamefont{A.}~\bibnamefont{Georges}},\ and\ \bibinfo {author}
  {\bibfnamefont{A.}~\bibnamefont{Toschi}},\ }%
  \bibfield{journal}{%
  \Doi{10.1103/PhysRevB.85.085124}{\bibinfo {journal} {Phys. Rev. B}}\ }%
  \textbf{\bibinfo {volume} {85}},\ \bibinfo {pages} {085124} (\bibinfo {year}
  {2012})%
  \bibAnnoteFile{NoStop}{Taranto:2012}%
\bibitem{Toschi:2012}%
  \BibitemOpen
  \bibfield{author}{%
  \bibinfo {author} {\bibfnamefont{A.}~\bibnamefont{Toschi}},\ }%
  \enquote{\bibinfo {title} {{Private Communication}},}\ %
  \bibAnnoteFile{NoStop}{Toschi:2012}%
\end{thebibliography}%

\end{document}